\documentclass{appolb}
\usepackage{graphicx}
\def\ga{\,\,\raise0.14em\hbox{$>$}\kern-0.76em\lower0.28em\hbox{$\sim$}\,\,}
\def\la{\,\,\raise0.14em\hbox{$<$}\kern-0.76em\lower0.28em\hbox{$\sim$}\,\,}

\begin{document}
\title{Recent shell-model calculations of $\gamma$-decay strength functions
\thanks{Presented at XXVI Nuclear Physics Workshop}%
}
\author{Kamila Sieja
\address{Universit\'e de Strasbourg, IPHC, 23 rue du Loess 67037 Strasbourg, France,
CNRS, UMR7178, 67037 Strasbourg, France}
\\
Stephane Goriely
\address{Institut d'Astronomie et d'Astrophysique, Universit\'e Libre de Bruxelles, CP-226, 1050 Brussels, Belgium}
}
\maketitle
\begin{abstract}
We present recent shell-model calculations of the $\gamma$-decay
in $sd-pf$ and $pf$-shell nuclei. We focus on the
$M1$ part of the dipole strength which was shown to exhibit interesting low-energy effects, 
in particular a low-energy enhancement which can have a considerable impact on the radiative neutron capture.
We discuss the persistence of the shell effects in the nuclear quasi-continuum
and the relation between the shape of the strength function at low energy and nuclear deformation.

\end{abstract}
  
\section{Introduction}

Theoretical calculations of the neutron capture cross sections within the Hauser-Feshbach 
model \cite{Hauser-Feshbach} require the knowledge of the $\gamma$-decay probability of the compound nucleus
which is characterized by a statistical de-excitation strength function. 
It is known that the photon strength function is dominated by the dipole component which has traditionally
been modeled by simple Lorentzian approximations with some energy dependence \cite{Capote2009}.
Such models however cannot describe structure effects at lowest $\gamma$ energies 
which were shown to have a considerable impact on the calculated neutron capture cross sections for exotic neutron-rich nuclei \cite{Larsen2010}.   
A systematic microscopic evaluation of dipole strength functions was achieved in the QRPA approach 
\cite{Martini2016, Goriely2016} however only for the photoabsorption strength.
In recent years dipole strength functions were also obtained within the large-scale shell model (SM)
in several regions of the nuclear chart \cite{Schwengner-Mo, Schwengner-Fe, Brown-fe56, Mitbo-ni70, Mitbo-LEE, Sieja-PRL, Sieja2018}, 
proving the capacity of this framework
to provide at least qualitative explanation of low-energy effects revealed by experimental data.
In particular, in Ref. \cite{Sieja-PRL} a first microscopic calculation of both dipole
modes was performed, showing a different behavior of the $E1$ and $M1$ strengths at low 
transition energy. The trends predicted by shell model were later incorporated 
to a semi-empirical, global description of $\gamma$-decay strengths for applications 
in Ref. \cite{Goriely2018}, showing an overall improvement of the calculated radiative widths and 
neutron capture cross sections as compared to those obtained using simple analytical prescriptions
for the de-excitation strength functions. It was also predicted that such an upbend of the $M1$
strength towards $E_{\gamma}=0$, if present, can considerably affect the calculated   
neutron capture rates. 

Shell-model studies suggested however that the enhancement of the low-energy $M1$
strength is most probable near closed shells, where protons and neutrons 
occupy high-$j$ orbitals of different parity and that the low-energy strength can be shifted 
to the scissors mode in deformed nuclei \cite{Schwengner-Fe}. A relation between the strength at low energy and 
$B(E2)$ values, being the measure of nuclear deformation, was presented also in Ref. \cite{Sieja2018}.
In the present work we explore further shell-model
systematic calculations in $pf$ and $sd-pf$ nuclei, which shed more light on the relation between the low-energy behavior 
of the $M1$ strength and the nuclear shape. 

\section{De-excitation strength function in the shell-model framework}
\label{sec2}

In the following, we present systematic calculations of the de-excitation $M1$
strength function in a number of isotopic chains that can be described 
in the $sd-pf$ and $pf$ model spaces. In the present calculations we treat 
the full model space and use well-established shell-model interactions
SDPF-U \cite{SDPF} and LNPS \cite{Lenzi2010}, respectively.
A standard quenching factor of 0.75 is applied to the spin part of the magnetic dipole operator.
The de-excitation strength function is obtained from the Bartholomew definition \cite{Bartholomew}:
\begin{equation}
f_{M1}(E_{\gamma}, E_i, J_i, \pi)=16\pi/(9\hbar c )^3 \langle B(M1)\rangle \rho(E_i,J_i, \pi),
\label{eq-fm1}
\end{equation}
where $\rho_i(E_i,J_i,\pi)$ is the partial level density determined at a given initial excitation energy $E_i$
and $\langle B(M1)\rangle$ the average decay probability for an energy interval.  
For each nucleus we calculate 60 excited states in the spin range $J=0-7$ for even and odd-odd nuclei and
$J=1/2-15/2$ for the odd ones.  
This typically leads to $\sim 2\cdot10^4$ of $M1$ matrix elements used in the averaging 
of $\langle B(M1)\rangle$ values in Eq.~\ref{eq-fm1}. Such a treatment 
was shown to provide strength functions which can be directly 
compared to experiment and give a qualitative explanation
of the latter \cite{Schwengner-Mo, Schwengner-Fe, Brown-fe56, Mitbo-ni70, Mitbo-LEE, Sieja-PRL, Sieja2018}.

\begin{figure}[htb] 
\centerline{\includegraphics[width=6.5cm]{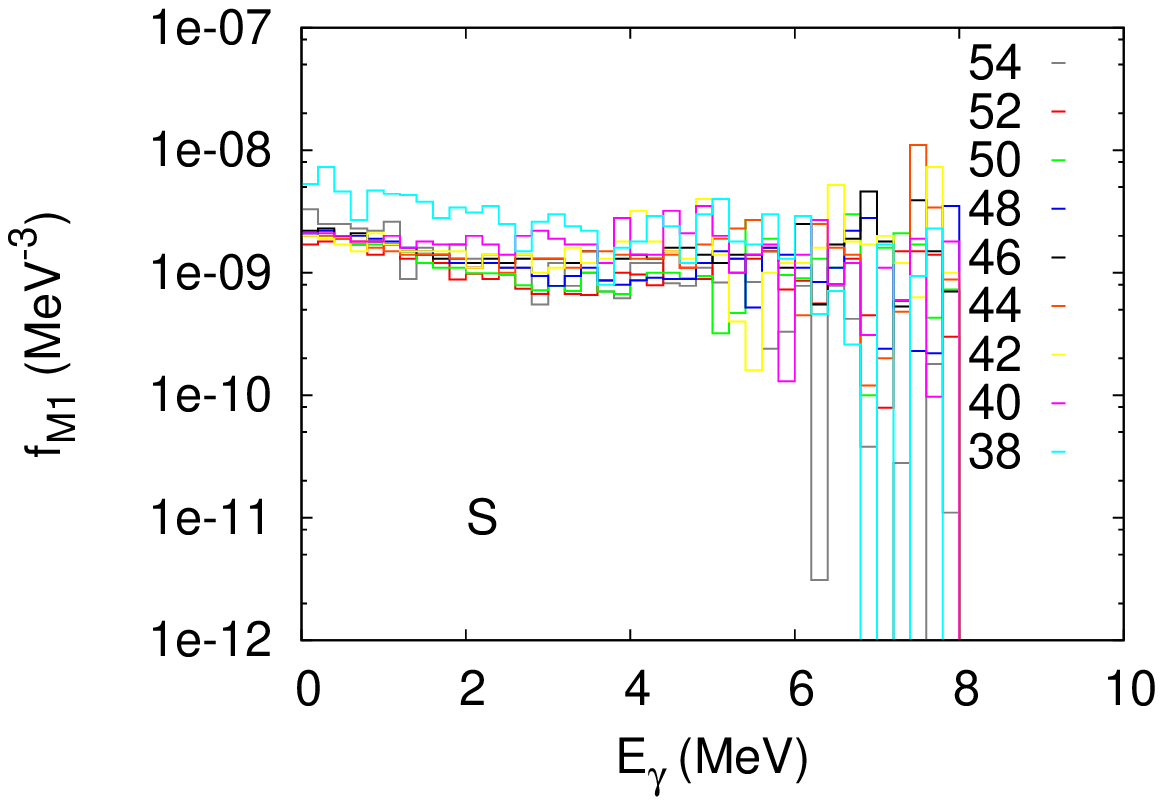}\includegraphics[width=6.5cm]{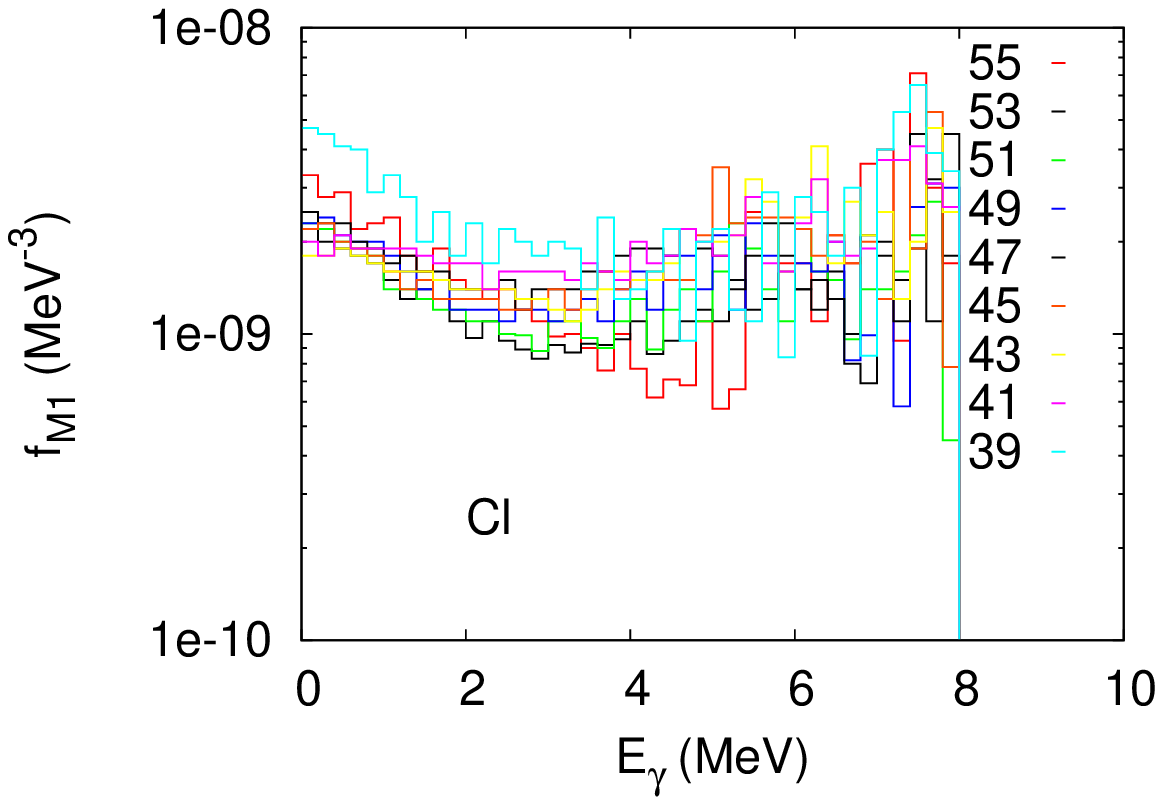}}
\centerline{\includegraphics[width=6.5cm]{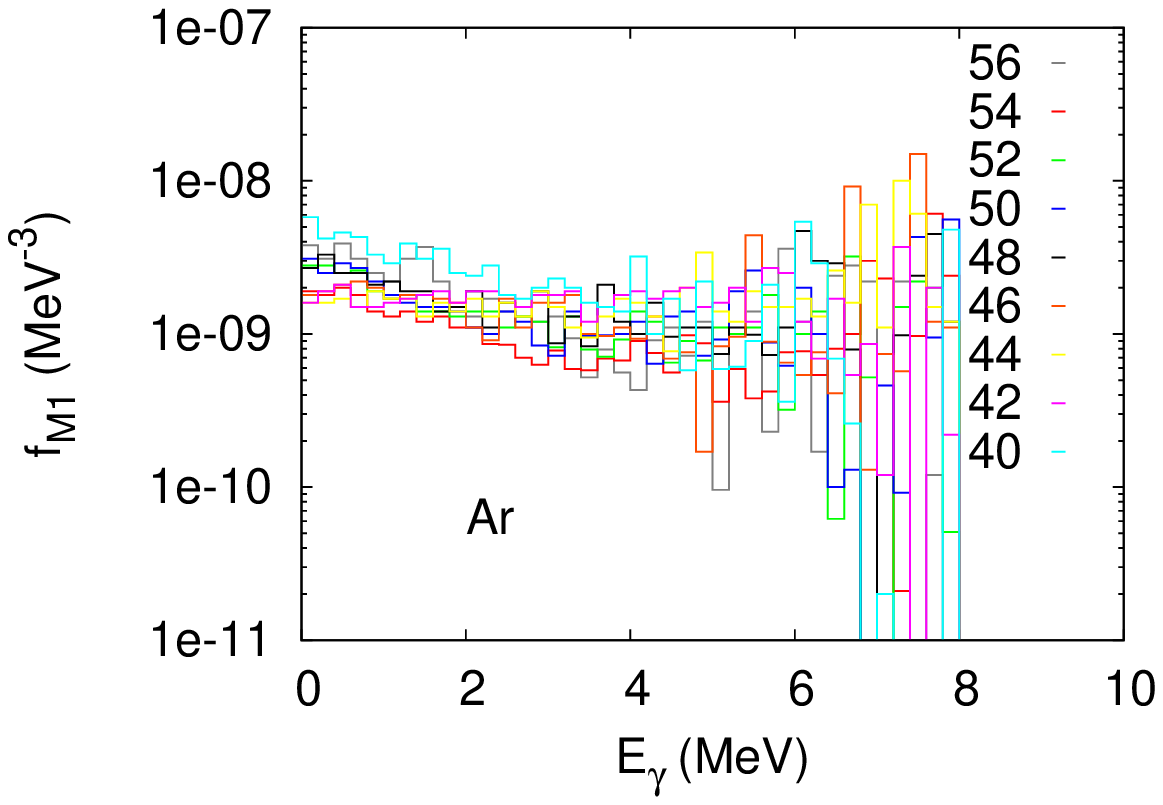}\includegraphics[width=6.5cm]{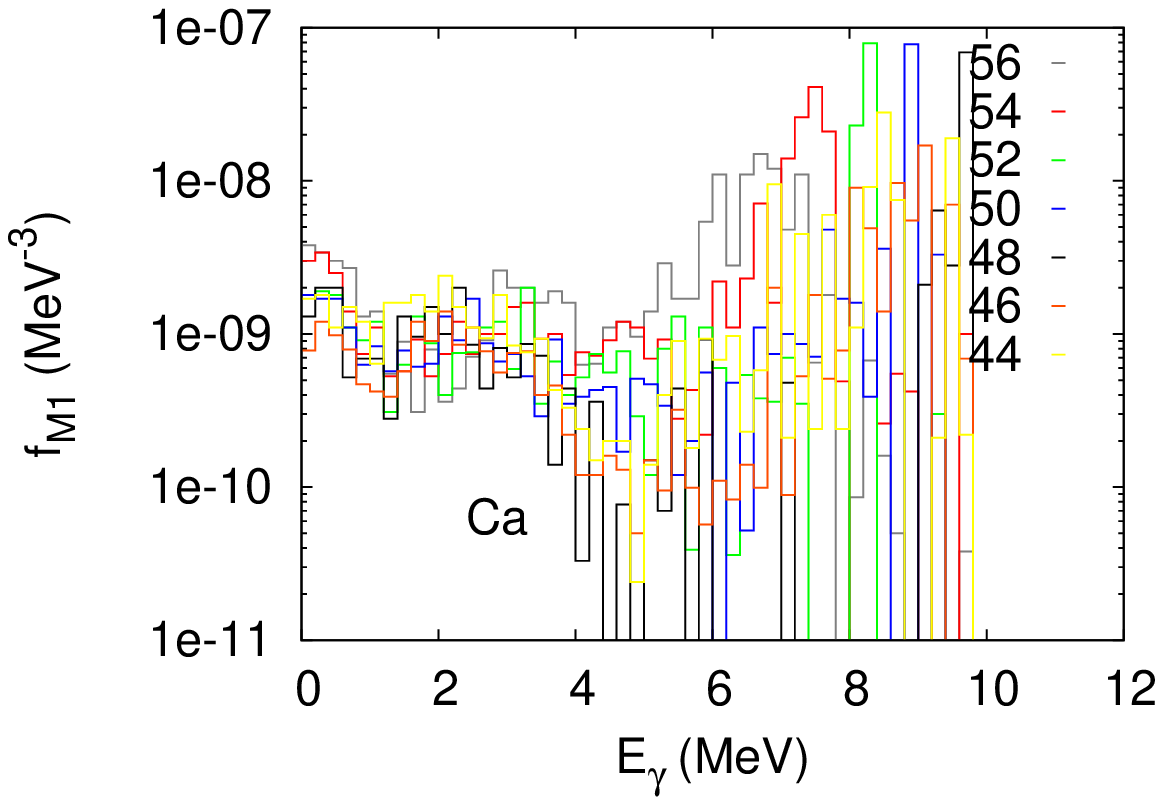}} 
\caption{De-excitation $M1$ strength functions obtained by shell-model calculations in several isotopic chains.}                              
\label{Fig:sf}                                                                                                                             
\end{figure}

In Fig.~\ref{Fig:sf}, we show the results of shell-model calculations in the 
even-even isotopes of sulfur, argon and calcium as well as in the odd-even
chlorine nuclei. As seen, the magnitude of the $f_{M1}$ is rather independent 
on the mass number for the lowest $\gamma$ energies. Slightly different trends
can be observed between  isotopic chains and between isotopes
in a given chain, as will be made more explicit in the next section. 

\begin{figure}[htb]
\centerline{\includegraphics[width=6.5cm]{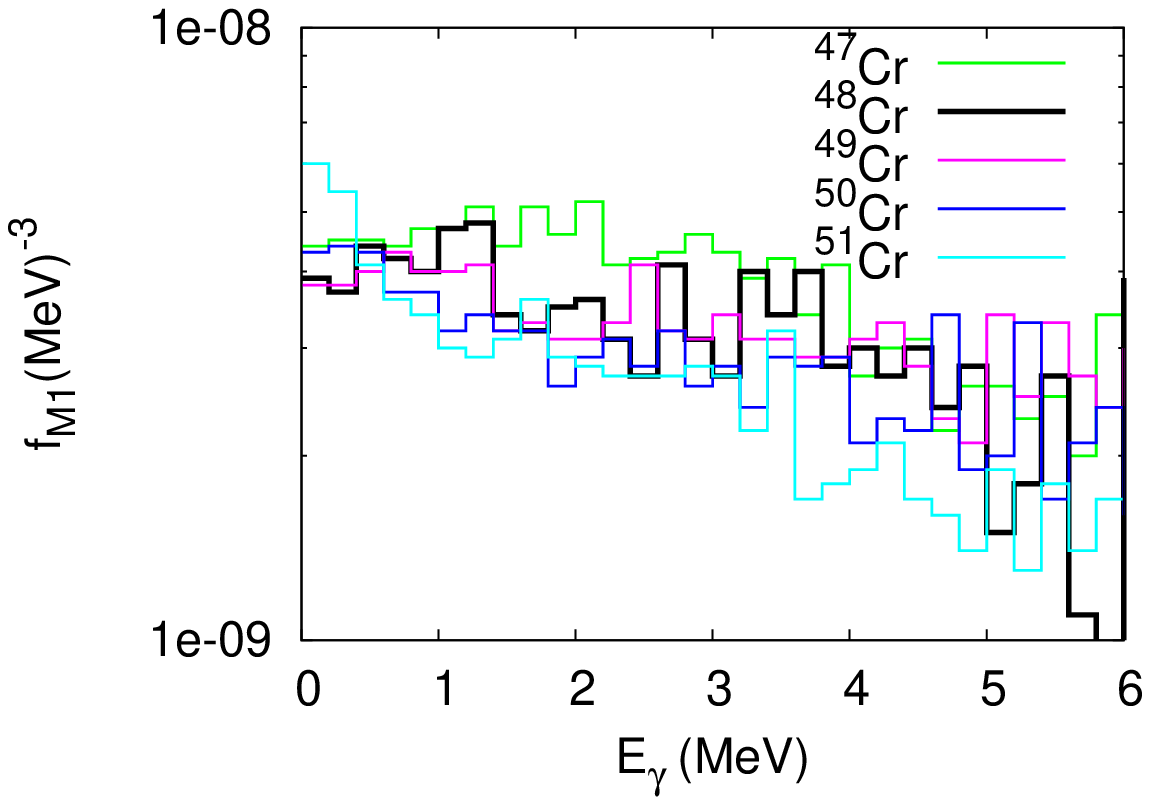}\includegraphics[width=6.5cm]{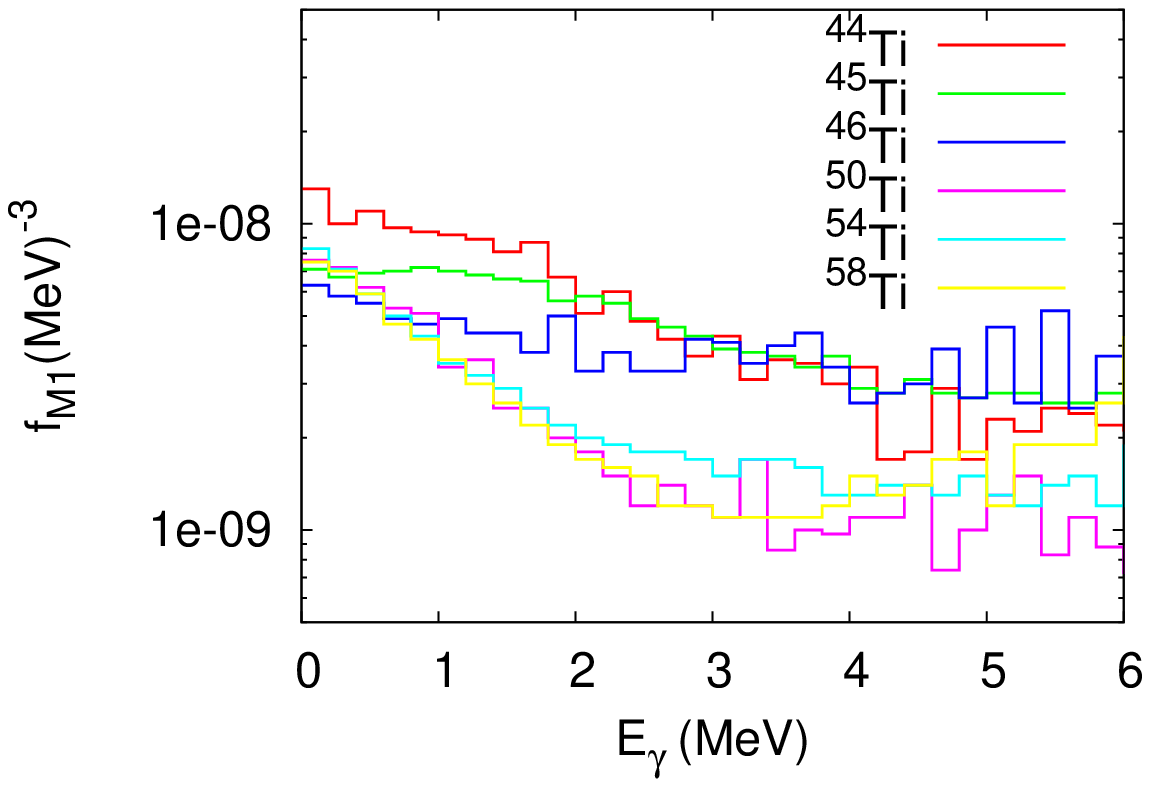}}
\caption{De-excitation $M1$ strength functions obtained by shell-model calculations in Cr and Ti chains.}
\label{Fig:sfpf}
\end{figure}

We also present, in Fig.~\ref{Fig:sfpf}, a systematic $M1$ calculation of
selected $pf$-shell nuclei, namely titaniums and chromiums. 
These two chains exhibit a different behavior: while nearly all the titanium isotopes 
have an upbend in their strength function, no such upbend is present in the chromiums,
up to $^{51}$Cr. One should stress that $^{48}$Cr is a model example of a prolate-deformed rotor in the $pf$-shell 
and its closest neighbours are also well deformed. 
As was stressed before, such deformed systems do not posses a spike 
at the lowest transition energy but a rather flat form of the $f_{M1}$ towards $E_{\gamma}=0$.

\section{Shell effects in the nuclear quasi-continuum}
The trends observed in the $\gamma$-ray strength function plotted in Figs. \ref{Fig:sf} and \ref{Fig:sfpf} 
are entirely due to the behavior of the averaged $B(M1)$ strength in the considered nuclei
(meaning the shape of $f_{M1}$ is independent of the calculated level density).  
As we stated above, the appearance of the upbend is related to the proximity of shell
closures and occupancies of specific neutron and proton orbitals. 
In particular, the authors of Ref. \cite{Mitbo-LEE} plotted a ratio of averaged $B(M1)$ values
in the energy intervals 0-2MeV and 2-6MeV, i.e. $\frac{\langle B(M1)\rangle (0-2MeV)}{\langle B(M1)\rangle (2-6MeV)}$.
Such a representation showed a clear correlation between the upbend and the proximity of a shell closure, i.e. 
the ratio in $sd$ and $pfg$ nuclei, with a few exceptions, exhibited a parabolic trend
typical of the $2^+$ excitation energies along the isotopic chains.

\begin{figure}[htb]
\centerline{\includegraphics[width=6.5cm]{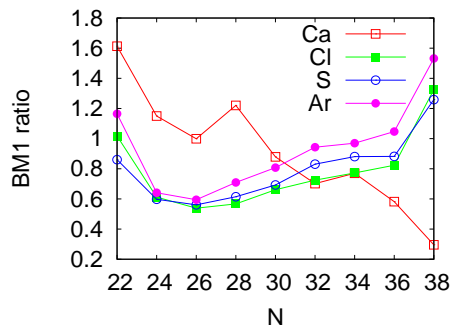}}
\caption{Ratio of the averaged $B(M1)$ strengths in calculated nuclei. See text for details.}
\label{Fig:ratio}
\end{figure}

In the following, we extract the ratio of the averaged $B(M1)$ strength
in the calculated isotopic chains from the $sd-pf$ shell, where the $N=20,28$ and $N=40$
shell closures should influence the behavior of the low-energy strength, see Fig. \ref{Fig:ratio}.

As can be noted, the ratio peaks towards the $N=20$ closure for all isotopic chains. 
The $N=40$ sub-closure is predicted to exist in S, Ar and Cl isotopes but disappears 
for the calcium chain. $^{60}$Ca was studied in Ref. \cite{Lenzi2010}
and suggested to have a considerable $4p-4h$ component in its ground state in accordance
with the present observation of no-shell effect in the $f_{M1}$.    
In contrast, the $N=28$ shell closure is visible only in calcium. This is consistent 
with the current evidence for the sulfur isotopes, where the shape-coexistence 
was found at $N=28$ \cite{Force-S44, Egido-S44}. On the contrary, 
the $N=28$ closure in $^{46}$Ar was predicted from mass measurements, $2^+$ energies
and from $B(E2)$ values, see e.g. \cite{AME2016, Calinescu-Ar46}. Interestingly, the present interaction reproduces well the first two
but not the $E2$ transition strength. It was shown in Ref. \cite{Calinescu-Ar46} that none of the available shell-model interactions
in the $sd-pf$ model space is able to reproduce the experimental value of $B(E2; 2^+\rightarrow0^+)$
in $^{46}$Ar. One can thus expect that a similar problem concerns the magnetic transition strengths
and that the de-excitation strength function may be influenced by the $N=28$ shell closure also 
in the argon chain.

The independence of the $\gamma$-ray strength function on excitation energy, which
was tested in the present and in many other shell-model calculations, means that
the shell effects are also independent on the excitation energy and survive 
close to the neutron threshold. Thus the $\gamma$-ray de-excitation strength function can be used
as another probe of the shell effects and of their persistency at higher excitation energies.

\section{Deformation dependence of the low-energy $M1$ limit}

\begin{figure}[htb]                                                                                                                         
\centerline{\includegraphics[scale=0.33]{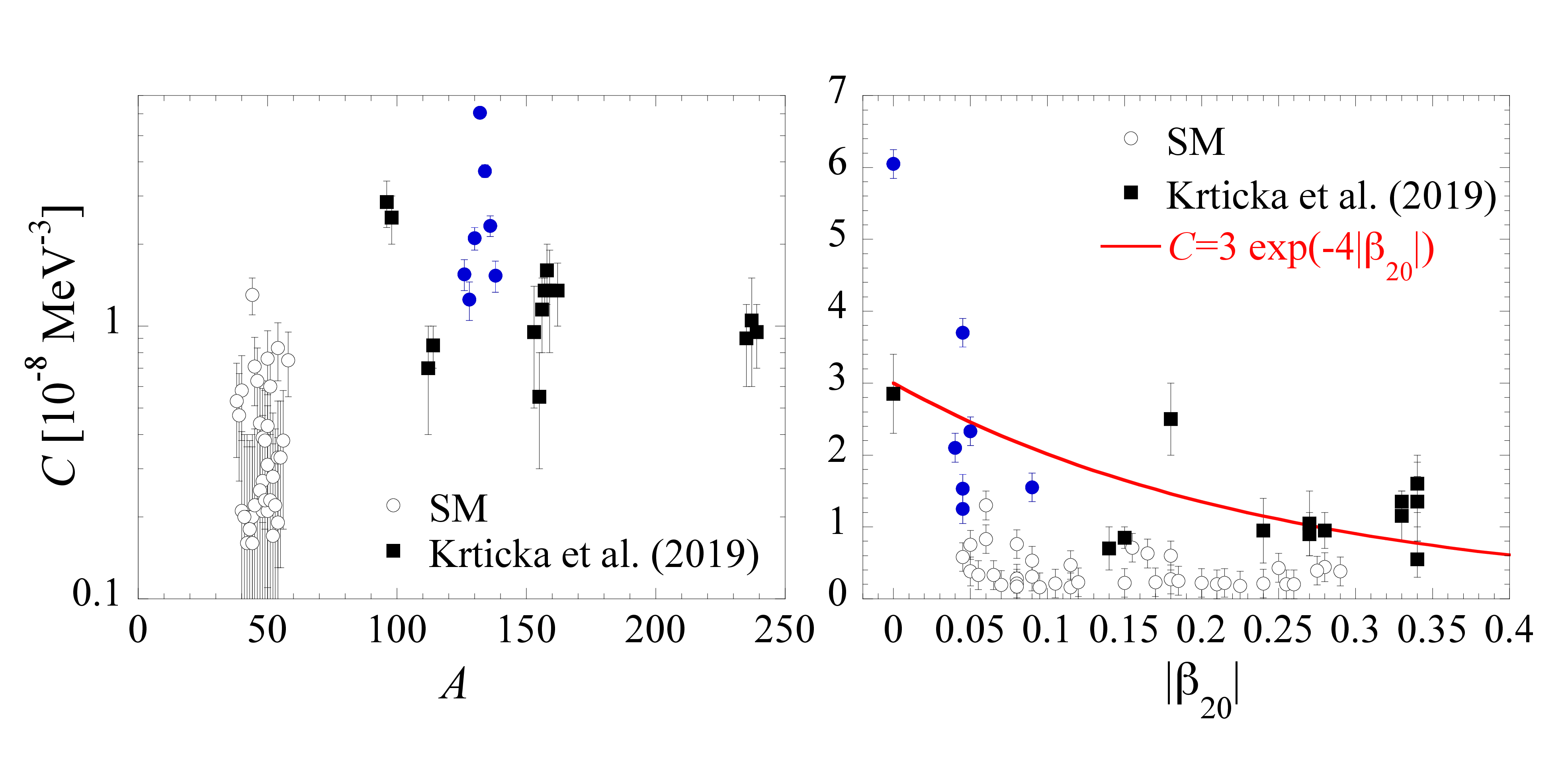} }
\caption{SM (circles) and experimental (squares) \cite{Krticka2019} estimates of the low-energy enhancement factor $C$ as a function of the atomic mass 
$A$ (left panel) of D1M deformation parameter $\beta_{20}$. The open black circles correspond to SM calculations presented in Sect.~\ref{sec2} 
and the blue solid circles to SM calculations for nuclei in the vicinity of $^{138}$Ba \cite{Sieja2018}. The red curve is the simple empirical 
energy-dependence adopted in Ref.~\cite{Goriely2019}. An error bar of $0.2 \times 10^{-8}$~MeV$^{-3}$ has been included in all SM estimates. }                                                                                                                       
\label{Fig:C}                                                                                                                             
\end{figure}

In our previous work \cite{Goriely2018}, we introduced an empirical low-energy contribution to the
$M1$ and $E1$ de-excitation strength function to complement the axially-symmetric QRPA predictions 
based on HFB calculations with the Gogny D1M interaction \cite{Martini2016, Goriely2016}. 
In particular, the so-called D1M+QRPA+0lim de-excitation $M1$ strength function was expressed as
\begin{equation}
\overleftarrow{f_{M1}}(E_\gamma) =  f_{M1}^{QRPA}(E_\gamma) +  C ~ e^{-\eta E_\gamma}  \label{eq2}
\end{equation}
where $f^{QRPA}_{M1}$ is the D1M+QRPA $M1$ strength function at the photon energy 
$E_\gamma$ and $C \simeq 10^{-8}$~MeV$^{-3}$, $\eta=0.8$~MeV$^{-1}$ are free parameters that were adjusted 
on shell-model results and available low-energy experimental data such as those obtained with the Oslo method \cite{Voinov,Guttormsen2005,Algin2008}. 
However, due to the lack of systematic calculations, the low-energy limit of the $M1$ strength was assumed  to be deformation independent. 
A first attempt to determine the deformation dependence of this $M1$ low-energy enhancement was done in Ref.~\cite{Krticka2019} on 
the basis of experimental data where constraints could be imposed from multistep $\gamma$-cascade spectra extracted from neutron capture on isolated resonances. 
The resulting experimentally constrained $C$ values are shown in Fig.~\ref{Fig:C}. Such multistep cascade data is available for a set of 15 nuclei only, mainly deformed, 
including isotopes of Mo, Cd, Gd, Dy and U. For this reason, it was proposed to adopt a lower limit, $C= 10^{-8}$~MeV$^{-3}$, for all nuclei with $A\ga 105$ and 
for lighter nuclei a simple deformation-dependent $C=3 \times 10^{-8} \exp(-4\beta_{20})$~MeV$^{-3}$ \cite{Krticka2019,Goriely2019}. The corresponding prescription 
is shown in Fig.~\ref{Fig:C} (right panel). With the present large-scale SM calculations, as presented in Sect.~\ref{sec2}, it is possible to test the relevance of such a prescription.

The SM and experimentally constrained  \cite{Krticka2019} estimates of the low-energy enhancement factor $C$ are summarized in Fig.~\ref{Fig:C} as a function 
of the atomic mass $A$ and of the deformation quadrupole parameter $\beta_{20}$. $C$ values lower than $10^{-8}$~MeV$^{-3}$ are obtained for light $sd$ and $pf$ 
nuclei. While non-negligible quadrupole deformation may be derived by mean-field models (see Fig.~\ref{Fig:C}, right panel), those nuclei are known as being vibrational, 
so that the D1M determination of their static deformation may not be the right description. In this case, the exponentially decreasing prescription as a function of the deformation, 
as proposed in Ref.~\cite{Krticka2019,Goriely2019}, is clearly not satisfactory.
Significantly larger $C$ values  between 1 and $6 \times 10^{-8}$~MeV$^{-3}$ are predicted by the SM for the spherical $N\simeq 82$ nuclei 
reaching this maximum value for $^{132}$Te \cite{Sieja2018}. A rather similar value was extracted from multistep $\gamma$-cascade spectra for the $^{96,98}$Mo isotopes.

The present comparison clearly shows that the simple prescription used so far, both as a function of $A$ and $\beta_{20}$, may not be 
adequate to describe the complex structure of this low-energy contribution to the $M1$ de-excitation strength function and that much more theoretical and experimental work is needed. 

\section{Acknowledgements}
SG is F.R.S.-FNRS research associate


\bibliography{../../../../BIB/kama}
\bibliographystyle{unsrt}

\end{document}